\documentclass[]{spie}  %>>> use for US letter paper
\usepackage{amsmath,amsfonts,amssymb}
\usepackage{graphicx}
\usepackage[colorlinks=true, allcolors=blue]{hyperref}
\usepackage{subcaption}

\title{MiraSOL: a DMD-based spectrograph for resolved solar spectroscopy
}

\author[a,b]{Christian Robles}
\author[a,b]{Suvrath Mahadevan}
\author[a,b]{Lawrence Ramsey}
\affil[a]{Department of Astronomy and Astrophysics, The Pennsylvania State University, 525 Davey Laboratory, 251 Pollock Road, University Park, PA, 16802, USA}
\affil[b]{Center for Exoplanets and Habitable Worlds, The Pennsylvania State University, 525 Davey Laboratory, 251 Pollock Road, University Park, PA, 16802, USA}

\authorinfo{Further author information: (Send correspondence to C.R.)\\C.R.: E-mail: crobles@psu.edu}

\begin{document} 
\maketitle

\begin{abstract}
We present the science motivation, preliminary design requirements, device laboratory testing and a prototype for an new experimental platform for solar observations, MiraSOL. MiraSOL will use digital micromirror technology to actively select regions on the solar disk for spectroscopic observation to determine the spatially dependent radial velocity signatures of stellar variability, and can create transits on the solar disk to probe the effects of stellar contamination on exoplanet transmission spectra. MiraSOL uses the Texas Instruments DLP801RE as a spatial light modulator to allow a mask, with 3 arcsecond spatial sampling per micromirror,  capable of resolving features on the solar disk. This instrument will have a fiber output which can then be coupled with state-of-the-art extreme precision radial velocity (EPRV) spectrometers, such as HPF or NEID, for high resolving power, stable spectra of sunspots and plage, or a low resolution spectrometer for studies of stellar contamination in transit spectra. We discuss  a 60 Hz flicker signal we discovered, likely due to the commercial off-the-shelf (COTS) evaluation board of the digital micromirror device electronics. We also build a proof-of-concept prototype and demonstrate imaging and pixel-level control of the full solar disk to demonstrate the feasibility of this technology.
\end{abstract}


\section{INTRODUCTION}
\label{sec:intro}
With the upcoming Habitable Worlds Observatory\cite{decadal}, there is a need to find habitable-zone exoplanets around the nearest Sun-like stars to target for reflected light spectroscopy. Due to the low geometric transit probability of these planets, the radial velocity method remains the best option for finding these nearby terrestrial worlds. While the extreme precision radial velocity (EPRV) community has been improving the {\it instrumental} precision of state-of-the-art instruments to approach 10 cm/s\cite{Schwab2016,Pepe2021}, stellar variability currently prevents the detection of smaller worlds\cite{Burt2025,Crass2021}. Observing the Sun-as-a-star with dedicated solar feeds attached to these EPRV spectrographs provides high-quality data sets to test stellar variability mitigation methods\cite{Ford2024} and monitors instrument performance relative to other EPRV spectrographs\cite{Zhao2023}. 

Current Sun-as-a-star instruments integrate the entire solar disk, which dilutes the signal of individual active regions. Spatially resolved radial velocities are necessary to isolate the localized effects of convective blueshift suppression and magnetic activity.
To bridge this gap, new experiments are being designed and conducted including PoET\cite{Santos2025}, ABORAS\cite{FarretJentink2022} and the Institute for Astrophysics Gottingen Fourier Transform Spectrograph\cite{Schafer2020} for spatially resolved observations on the Sun.
We are constructing MiraSOL (MIcromirrors for Resolved Adaptable Solar Observations Limb-to-limb) to address this need for more advanced solar feeds that can dynamically view the solar surface.
In Sec.~\ref{sec:req} we present the key design requirements for MiraSOL, followed by a discussion of laboratory testing of the High Efficiency Pixel Digital Micromirror Devices (DMDs) in Sec.~\ref{sec:dmd}. In Sec.~\ref{sec:schematic} we discuss the instrument schematic and future plans for the MiraSOL instrument. In Sec.~\ref{sec:proto} we build and test a MiraSOL prototype to validate the application of DMDs to solar observations.

\section{Design Requirements}
\label{sec:req}
\begin{figure}[h]
    \centering
    \includegraphics[width=16cm]{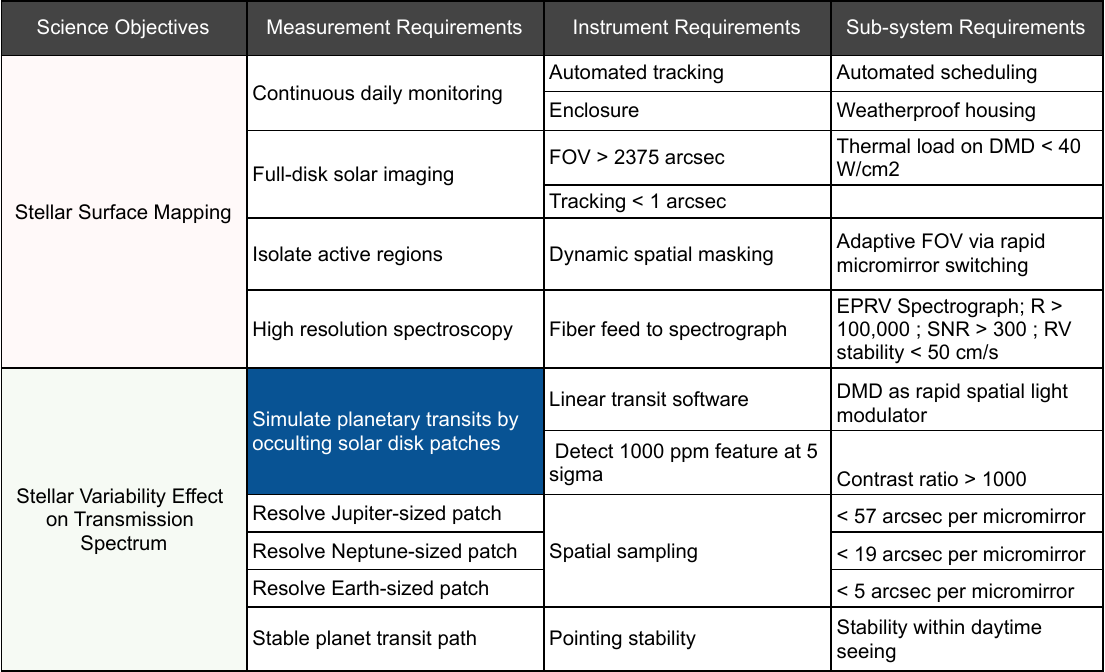}
    \caption{The Science Traceability Matrix for the MiraSOL instrument traces instrument requirements from the key science objectives for stellar surface mapping and studying the transit light source effect. The facility requirement of a fiber-fed EPRV spectrograph is included in the sub-system requirement column for matrix conciseness (though an EPRV spectrograph is an entire facility, not a sub-system). In practice, MiraSOL can be integrated with existing EPRV spectrometers like NEID\cite{Schwab2016}, HPF\cite{Mahadevan2012}, and EXPRES\cite{Jurgenson2016}.}
    \label{fig:stm}
\end{figure}

To specify instrument requirements, we first refined our science goals. Recent work on the convective blueshift from the Rossiter-McLaughlin\cite{Rossiter1924,McLaughlin1924} (RM) effect during Sun-as-a-star eclipse observations\cite{Reiners2016}, along with a lack of high-quality plage spectra for stellar variability models\cite{Cristo2025,Barka2026}, drive our surface mapping requirements. The modularity of the digital micromirror device (DMD) also allows us to explore the transit light source effect\cite{Rackham2019}, which arises from transmission spectra when a planet transits a star with surface inhomogeneities. We trace the measurement and instrument requirements from these science cases in our preliminary science traceability matrix (Fig.~\ref{fig:stm}). The following subsections detail these science cases and the flow-down of instrument requirements.

\subsection{Surface Mapping and Rossiter-McLaughlin Effect}
% science motivation
Stars vary in both intensity and radial velocity (RV). Understanding and mitigating this RV signal is essential for discovering terrestrial-mass exoplanets via the radial velocity method. This solar RV signal varies on timescales from years (solar cycles) to minutes (granulation)\cite{Cegla2019}. The radiative core and convective layer give rise to numerous magneto-convective features that appear as spots, plage, and granulation on the surface. Granulation is characterized by hot upwelling plasma (granules) surrounded by cool intergranular lanes of sinking plasma\cite{Gray2009}. The Doppler shift of the upwelling plasma dominates this signal, causing a bulk convective blueshift across the solar surface. The Sun's rotation also contributes a blueshift on the approaching limb and a redshift on the receding limb. These combined effects are visible during the RM effect, where an occulting body transits the solar disk, differentially blocking the RV signal from a specific patch and enhancing the net blue or redshift from the unocculted side of the rotating star. Fitting models to this effect places limits on the bulk convective blueshift, which is normally inaccessible\cite{Palumbo2024}. If portions of the solar disk can be selectively blocked using a spatial light modulator (SLM), this bulk convective blueshift from granulation can be directly probed. Other surface features with distinct RV signals include sunspots and faculae/plage\cite{Haywood2016}. While solar physicists have narrow-bandpass spectra of these regions, obtaining high-resolution spectra over a broad bandpass would be highly valuable. These features also alter the overall stellar intensity, which directly impacts the transit light source effect\cite{Rackham2019}.

To monitor the Sun and map its surface, the instrument requires daily observations, consistent with current Sun-as-a-star feeds. This necessitates a weatherproof design for roof installation without daily deployment, similar to the robust housing of the NEID solar telescope\cite{Lin2022} or the acrylic dome of the HARPS-N solar telescope\cite{Dumusque2021}.
The instrument also requires automated software to track the Sun throughout the day and effectively control the static states of the micromirrors. While MiraSOL will be multi-functional and run various operational scripts, it must also be capable of capturing standard Sun-as-a-star observations in line with other solar feeds.
To fully image the Sun on the DMD, the field of view (FOV) must be at least 2375 arcsec. Given the solar angular diameter of roughly 1900 arcsec, a 25\% margin is necessary for background subtraction and tracking errors. Because the instrument maps the solar image onto the DMD to control specific spatial regions, knowing the exact position of the Sun on the micromirrors is critical for accurate surface mapping.

To obtain spectra from specific active regions, the DMD provides adaptive spatial selection to the spectrograph, while the imaging arm provides a simultaneous image of the Sun with the selected regions appearing dark. This dual-arm approach ensures accurate spatial selection by generating a map that verifies the exact locations of the extracted spectra.
% high res spectra
For stable, high-resolution spectroscopy, MiraSOL will be paired with an EPRV spectrograph such as NEID in the optical or HPF in the NIR. The NASA EPRV working group\cite{Crass2021} recommends that solar feeds utilize spectrographs with better than 50 cm/s stability, a spectral resolution of $R > 100,000$, and a signal-to-noise ratio of SNR $> 300$ in the visible. While HPF does not meet these specific visible-light requirements, its NIR coverage complements visible solar feeds, bridging observations to longer wavelengths.

\subsection{Stellar Contamination on Transmission Spectra}
To study the transit light source effect, the DMD acts as a rapid spatial light modulator. By switching specific micromirrors to the OFF state, the solar signal from that spatial region is removed. Rapidly updating these micromirror states simulates a bare-rock planet transiting the solar disk. The resulting spectra provide a ground truth to better study the transit light source effect. To enable this, the DMD must switch faster than typical exposure times to smoothly track the simulated transit. The High Efficiency Pixel (HEP) DMDs operate at 120 Hz, which is more than sufficient for exposures lasting seconds or minutes. No current instrument has the capability to produce a simulated bare-rock transit directly across the resolved solar disk.

Generating these transits requires custom software to dynamically switch the micromirrors. The software must be flexible enough to produce transits of varying sizes, durations, and chords across the disk.
The simulated transit depth depends on the area of the occulted region. A Jupiter-sized planet around a Sun-twin produces a transit depth of roughly 1\%. Simulating a Jupiter-sized transit is desirable because the larger signal is easier to detect and induces a larger stellar contamination effect. To simulate smaller planets (e.g., Neptune or Earth) or to exercise precise shape control, the simulated planet must span multiple micromirrors. Consequently, we calculate our spatial requirements assuming a minimum $3 \times 3$ micromirror region to define the transit size.

\subsection{Oblate Planets}
Highly precise transit observations have opened the door to studying planetary oblateness by examining ingress and egress anomalies. A Jupiter-sized planet with Saturn-like oblateness produces a deviation of approximately 200 ppm compared to a spherical planet\cite{Liu2024}. Software packages currently simulate this oblate effect for observatories like James Webb Space Telescope; a ground-truth bare-rock oblate transit across the solar disk could validate and refine these models\cite{Dholakia2024}. Oblate transit models also provide constraints on planetary rotation periods, which in turn inform formation and spin-down mechanisms\cite{Lammers2024}. An instrument capable of altering the shape of the occulting feature, not just the transit depth, can test these models empirically.

%To test the effect of oblateness on the transit curves, sufficient obscuration shape control is needed to create non circular transit shapes. One micromirror would create a square while more could create more complex shapes. For a given transit depth the image size of the solar disk on the DMD will decide how many micromirrors are required to form the simulated planet. If the image on the DMD is too small then sufficient shape control will not be exercised. The DMD can not be illuminated outside its active area of micromirrors because doing so will heat up the DMD chip and could lead to permanent damage. To prevent this there is an artificial boundary made of 20 micromirrors in every direction coined the {\it pond of mirrors} to establish a buffer zone. To create Saturn like oblateness one axis of the planet needs to be 90\% of the other. To create this oblateness a pixelated circle of 10 pixels across and 9 pixels high could be created. This would require 80-76 pixels to represent Saturn depending on the circle shape. Further work into the generation of oblate pixelated planets is needed to better define instrument requirements.

To test the effect of oblateness on transit light curves, the DMD must provide sufficient shape control to create non-circular transit profiles. A single micromirror creates a square, whereas a larger array can approximate complex shapes. For a given transit depth, the size of the solar image on the DMD dictates how many micromirrors are available to form the simulated planet. If the solar image is too small, sufficient shape control cannot be achieved. 

Furthermore, the DMD cannot be illuminated outside its active micromirror area, as this causes heating and potential permanent damage to the chip. To prevent this, an artificial boundary of 20 micromirrors in every direction, coined the {\it pond of mirrors}, establishes a safe buffer zone. To simulate Saturn-like oblateness (where one axis is 90\% of the other), a pixelated ellipse approximately 10 pixels wide by 9 pixels high is required, utilizing roughly 76 to 80 pixels depending on the exact rasterization. Further work on the generation of oblate pixelated planets will refine these specific instrument requirements.

\section{DMD Evaluation}
\label{sec:dmd}

Our specific DMD device was chosen to optimize spatial fidelity, array size, and maximum illumination power. These design requirements led us to select the High Efficiency Pixel (HEP) DMDs, which feature filled central vias (artifacts in the center of micromirrors from manufacturing) unlike other DMD devices used in previous astronomical instruments such as SAMOS \cite{Smee2018}. Among the HEP models, we selected the DLP801RE; it offers a larger micromirror array than the DLP780RE and a higher peak illumination threshold than the DLP800RE, which is critical for solar imaging. While we believe this DMD is the best commercial option for our application, being an early adopter of this model presents unique challenges. 

\begin{figure}[h]
    \centering
    \includegraphics[width=14cm]{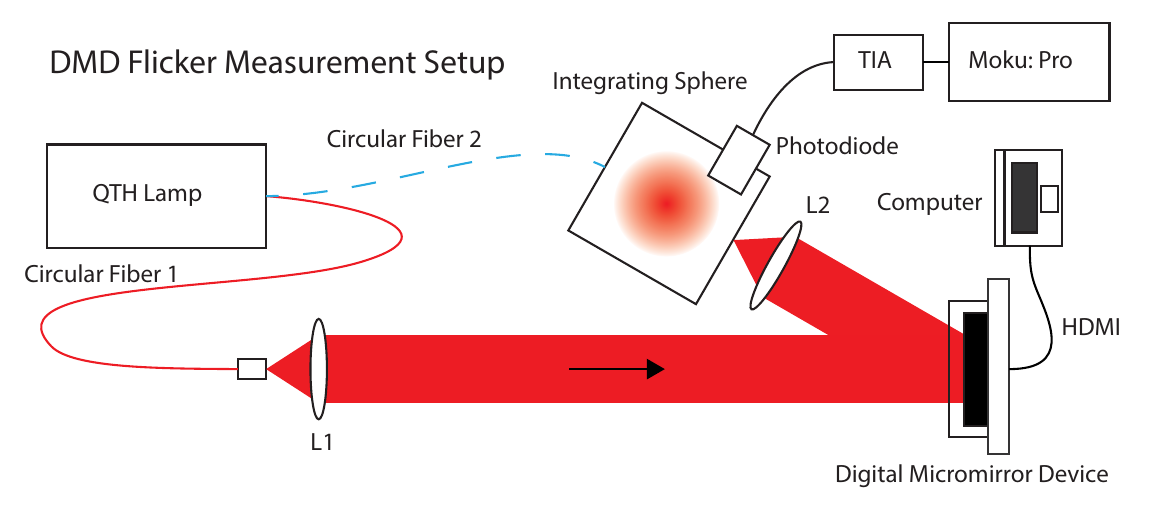}
    \caption{A schematic of our DMD testbench where we incorporate a Moku:Pro for kHz sampling of the optical signal to test the cause of the lower measured DMD contrast. Circular fiber 1 is used first to test the signal in the DMD path, and afterwards circular fiber 2 is used to bypass the DMD and directly illuminate the integrating sphere.}
    \label{fig:moku}
\end{figure}

The HEP DMDs and their associated Texas Instruments (TI) evaluation boards were designed and optimized for digital projectors \cite{Dewa2024}. By repurposing this DMD as a spatial light modulator for an external light source, we encountered unexpected behaviors when using the TI display-oriented evaluation electronics for scientific applications. 
To investigate the lower than expected functional contrast (250:1) reported in Robles \& Mahadevan (2026)\cite{Robles2026}, we modified our DMD testbench to enable faster sample acquisition while maintaining photometric precision. The baseline testbench uses a quartz tungsten-halogen (QTH) lamp (Thorlabs SLS201L) to illuminate the DLP801RE DMD. Light from the ON state is collected via a Hastings triplet (Thorlabs TRH254-040-A-ML) into an integrating sphere (Thorlabs 2P3). While our previous setup relied on a silicon photodiode (Thorlabs SM05PD1A) paired with a Keithley 6514 electrometer, this configuration lacked the temporal resolution needed to capture rapid micromirror flipping.
To measure the high frequency optical signal, we added a second silicon photodiode (Thorlabs SM05PD1A) to a port on the integrating sphere, as shown in Fig.~\ref{fig:moku}. The signal from this photodiode was amplified by a Femto DHCPA-100 transimpedance amplifier with a gain of $10^5$ V/A. The resulting voltage was recorded using a Liquid Instruments Moku:Pro. We utilized the Moku:Pro's DataLogger app, sampling at 10 kHz in precision mode with a 300 MHz bandwidth, 50 $\Omega$ input impedance, and a 4 V peak-to-peak range. We also used an alternative configuration in which the DMD is bypassed by directly injecting the light from the QTH into the integrating sphere.

\begin{figure*}[h]
    \centering
    \begin{subfigure}[t]{0.45\textwidth}
        \centering
        \includegraphics[width=\linewidth]{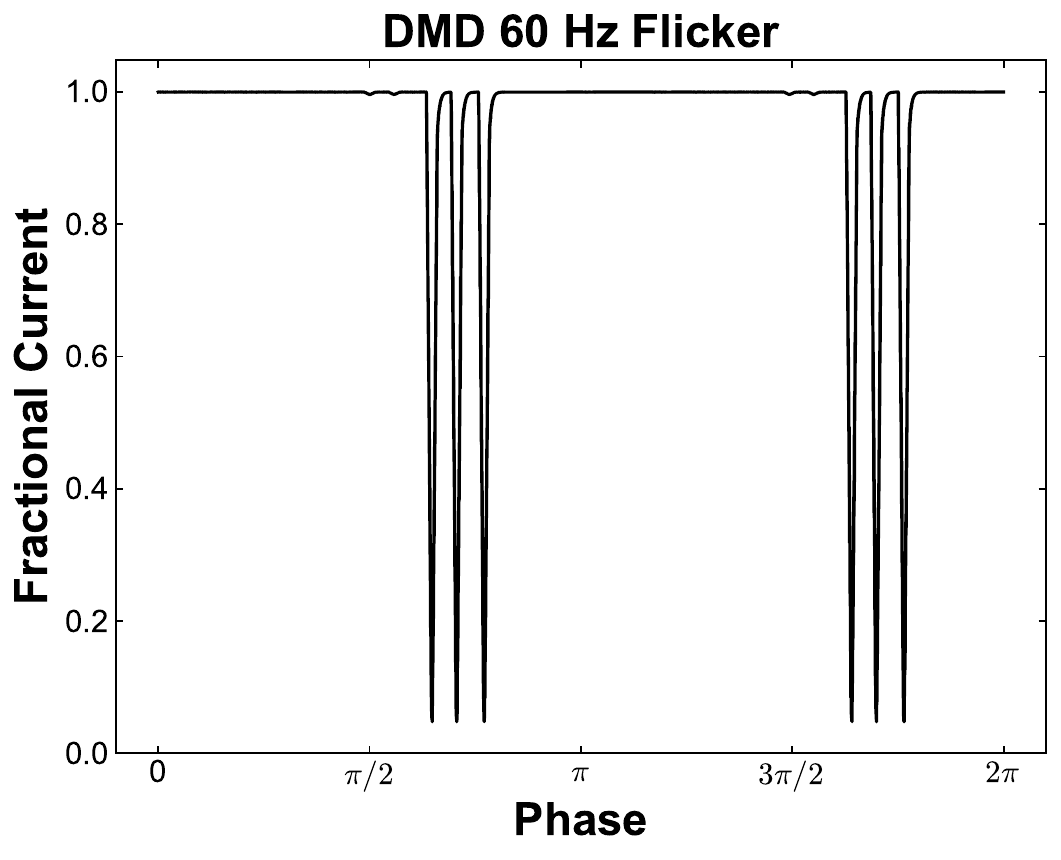} 
        \caption{} \label{fig:timing1}
    \end{subfigure}
    \hfill
    \begin{subfigure}[t]{0.45\textwidth}
        \centering
        \includegraphics[width=\linewidth]{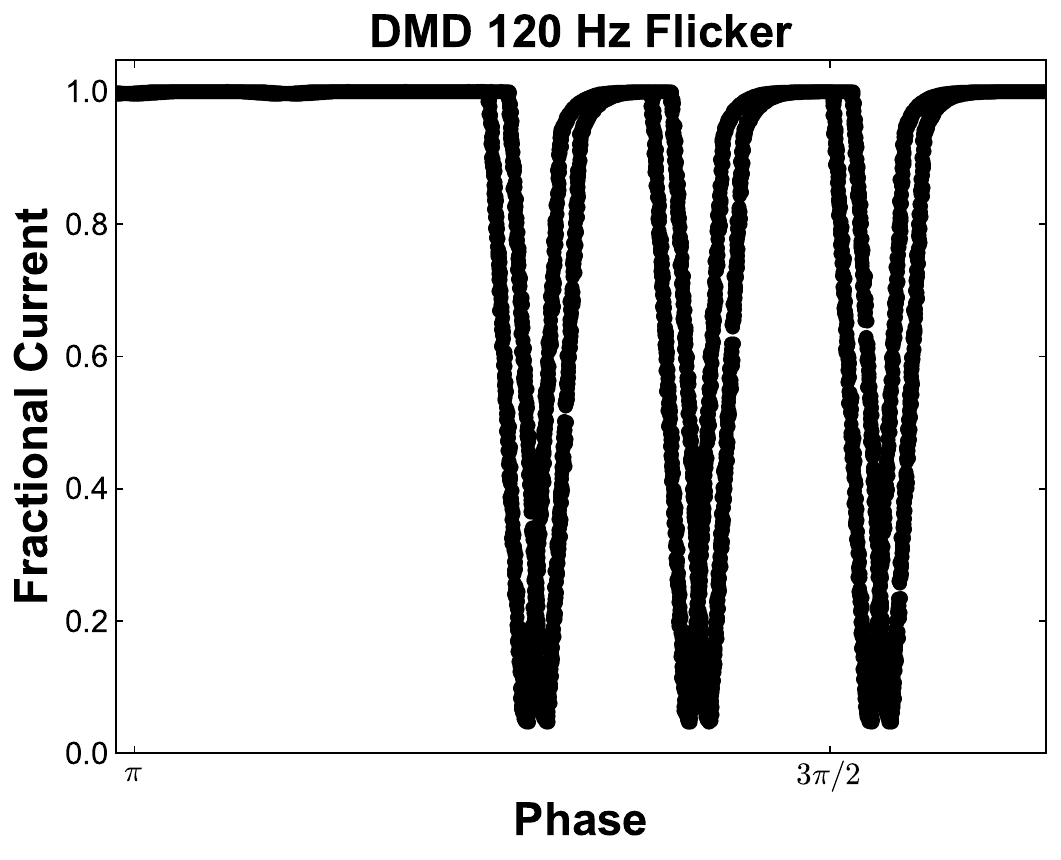} 
        \caption{} \label{fig:timing2}
    \end{subfigure}
    \caption{Using the Moku:Pro, the photocurrent is measured from the DLP801RE when displaying all ON state. (a) The signal, when phase folded to 60 Hz, shows two groups of three dips. (b) The 120 Hz phase-folded signal reveals that the two groups of three dips have a phase offset, indicating two distinct 60 Hz reset signals.}
\end{figure*}

Using this high speed setup, we measured the photocurrent from the DLP801RE while it was commanded to display an all white (all ON) state. We detected a persistent flickering signal introduced by the TI evaluation board, which likely drives the degraded contrast measured previously. Initially, this appeared as a 120 Hz signal; however, phase-folding the data revealed a phase shift between two halves of the signal, indicating two offset 60 Hz signals (Fig.~\ref{fig:timing1} and Fig.~\ref{fig:timing2}). We suspect this flicker is an automated micromirror reset sequence programmed into the TI evaluation board firmware to prevent micromirror stiction (hinge memory) during extended ON or OFF states.

\begin{figure}[h]
    \centering
    \includegraphics[width=16cm]{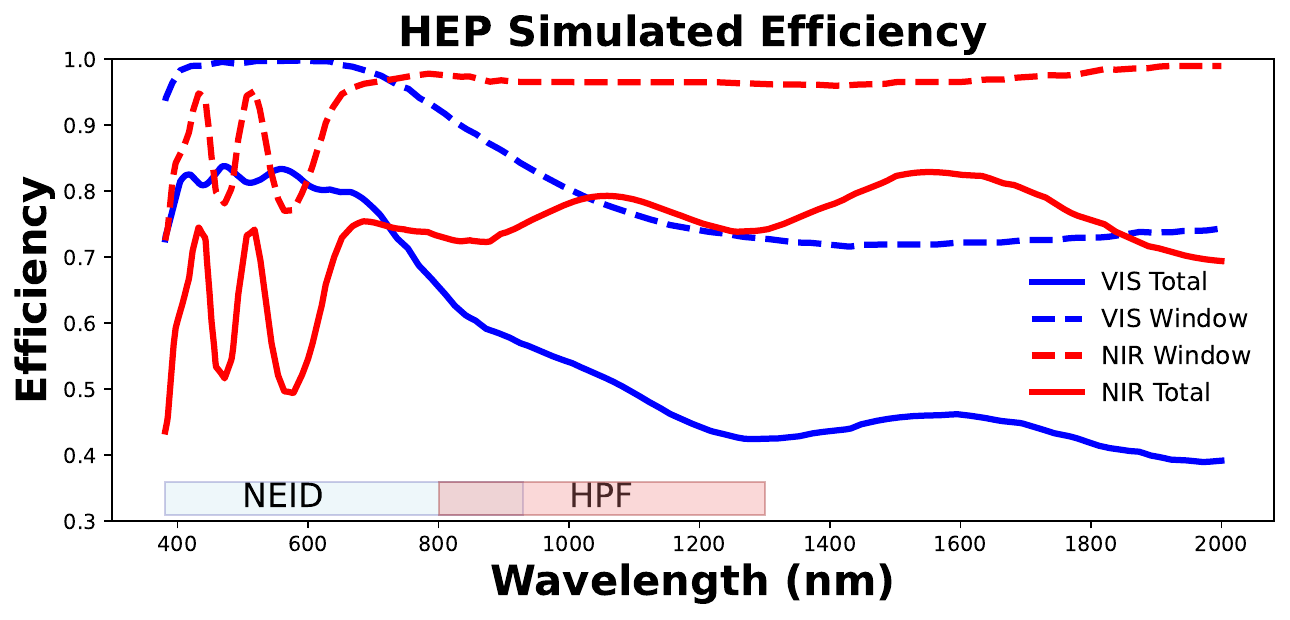}
    \caption{The NIR performance of the HEP DMDs is currently limited by the visible AR coating on the window. We simulate the efficiency of a hypothetical HEP DMD with a NIR coating, demonstrating that it would perform significantly better in the NIR and be better suited for use with the HPF spectrograph.}
    \label{fig:eff}
\end{figure}

Beyond timing electronics, we also simulated the DMD's optical efficiency. The stock window of the DLP801RE is optimized for visible wavelengths. However, if a Corning NIR optimized window were used, similar to options available for other DMD models, the optical efficiency would improve significantly in the NIR, better matching the wavelength range of the HPF spectrograph. Fig.~\ref{fig:eff} compares the simulated efficiency of the current visible-coated window against a hypothetical NIR-coated configuration with diffraction efficiency from Robles \& Mahadevan (2026)\cite{Robles2026}.

\section{Design Schematic}
\label{sec:schematic}
\begin{figure}[h]
    \centering
    \includegraphics[width=16cm]{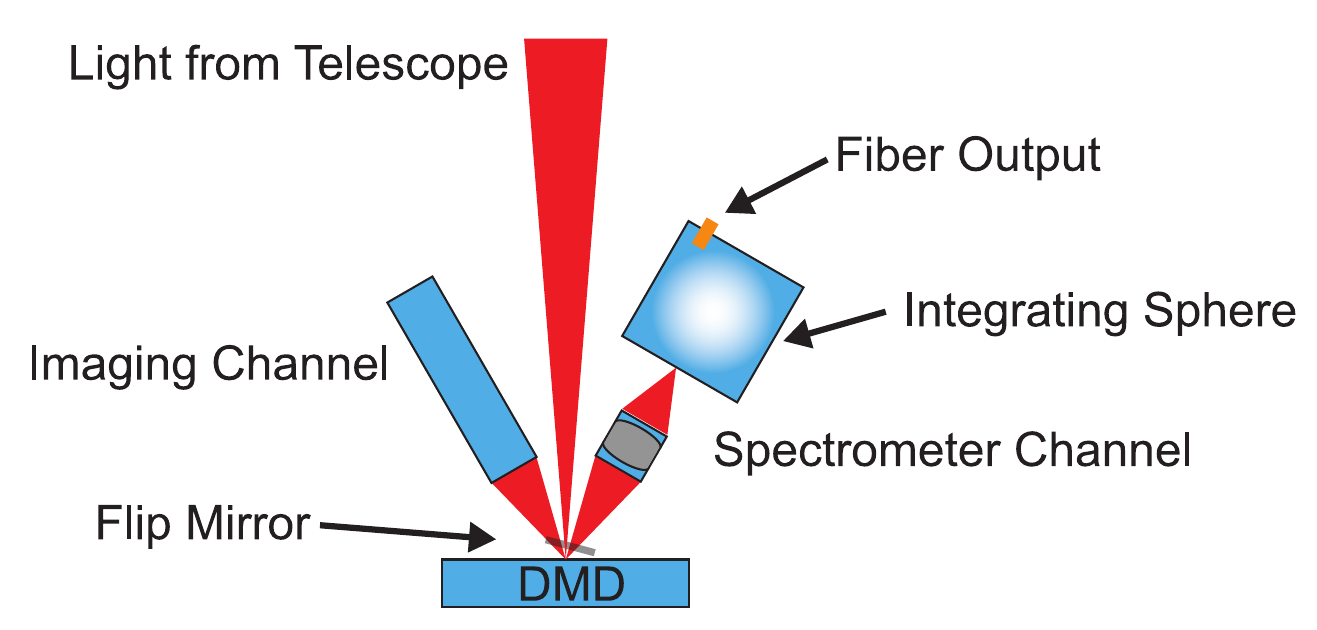}
    \caption{Preliminary optical schematic of the MiraSOL instrument. The DMD acts as a dynamic spatial light modulator, splitting the incident solar light into two distinct paths: a spectrometer channel (fed by the micromirror ON state) and an imaging channel (fed by the OFF state). A deployable fold mirror is also included to bypass the DMD for traditional disk-integrated Sun-as-a-star observations.}
    \label{fig:sch}
\end{figure}

The preliminary schematic of the MiraSOL design (Fig.~\ref{fig:sch}) illustrates how the DMD splits the incident solar light into two distinct paths: an imaging channel and a spectrometer channel. The solar disk is first imaged onto the DMD, which utilizes its individual micromirrors for dynamic spatial mapping. Each micromirror can be independently toggled to direct light into either of the two channels. 

Light directed to the spectrometer channel corresponds to the ON state of the micromirrors. This state was specifically chosen for the science channel to avoid scattered light from the inactive {\it pond of mirrors} surrounding the active area. The spectrometer channel will utilize imaging optics (currently under optimization) to couple the reflected light directly into the open port of an integrating sphere. The integrating sphere's fiber output can then be routed to a fiber-fed spectrograph. For stellar surface mapping, an EPRV spectrograph such as NEID or HPF would be used, whereas a lower resolving power spectrograph could be swapped in for transmission spectroscopy studies.
Conversely, light directed to the OFF state feeds the imaging channel, which is primarily used for target acquisition and simultaneous monitoring. This dual-channel approach provides a real-time image of the solar disk where the regions sent to the spectrograph appear dark. This unique capability allows us to visually confirm the outlines of sunspots or other targeted features, validating the exact spatial origin of the produced spectra. 
Additionally, we have included a deployable fold mirror that creates a direct light path to the integrating sphere. This bypass allows the instrument to function as a traditional, disk-integrated Sun-as-a-star feed when spatial modulation is not required. As the design matures, additional fold mirrors may be incorporated to condense the overall optical footprint.

\begin{figure*}[h]
    \centering
    \begin{subfigure}[t]{0.45\textwidth}
        \centering
        \includegraphics[height=10 cm]{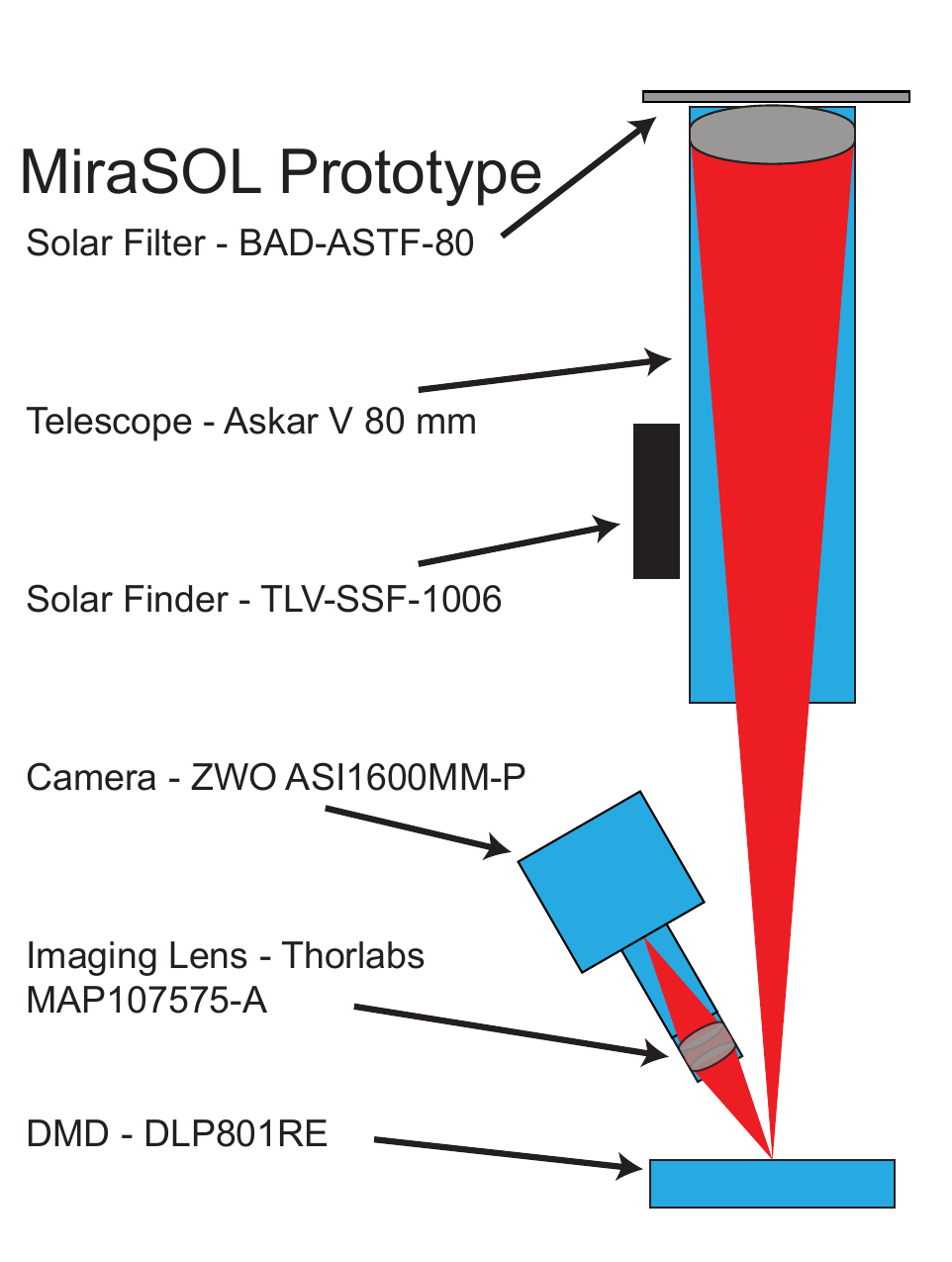} 
        \caption{} \label{fig:prototype_sch}
    \end{subfigure}
    \hfill
    \begin{subfigure}[t]{0.45\textwidth}
        \centering
        \includegraphics[height=10 cm]{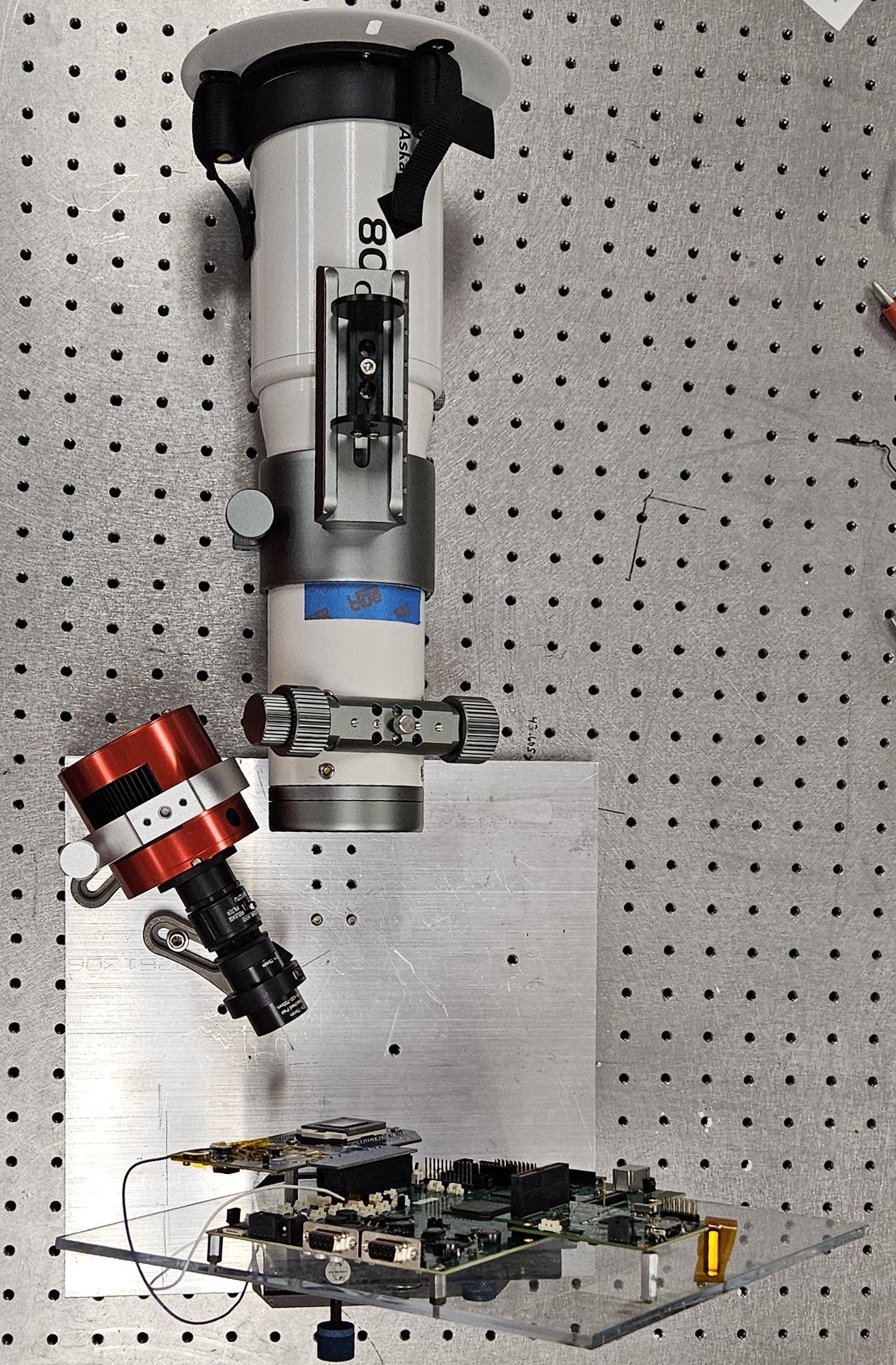} 
        \caption{} \label{fig:prototype_pic}
    \end{subfigure}
    \caption{Overview of the MiraSOL proof-of-concept prototype. (a) An optical layout and list of key commercial off-the-shelf (COTS) components used to validate the application of a DMD for spatially resolved solar observations. (b) A photograph of the assembled prototype with its protective light box removed, revealing the internal optomechanical layout and the imaging channel.}
    \label{fig:sch2}
\end{figure*}

\section{Prototype}
\label{sec:proto}
To establish the viability of using a DMD for solar observations, we built a proof-of-concept prototype. An overview of the key components and the optical layout are shown in Fig.~\ref{fig:prototype_sch}. We selected the Askar V telescope for its modularity, which features two aperture options (60 mm and 80 mm) and three interchangeable back lens groups (extender, reducer, and field flattener). To accommodate tight space constraints and eliminate the need for additional fold mirrors, our prototype utilizes only the 80 mm front lens group without any rear optics. A front-mounted Baader solar filter was used to attenuate the incoming light and prevent thermal damage to the DMD electronics.

This configuration yields a 500 mm focal length at f/6.25, producing a solar image with a diameter of roughly 511 micromirrors on the DMD surface. For our initial on-sky testing, we populated only the imaging channel using a ZWO ASI1600M-P. This setup allows us to visually verify the DMD's spatial selection capabilities across the solar disk. While the spectrographic performance of this specific DMD model has yet to be tested on-sky, its expected performance is simulated in Robles \& Mahadevan (2026)\cite{Robles2026}.

\begin{figure*}[h]
    \centering
    \begin{subfigure}[t]{0.45\textwidth}
        \centering
        \includegraphics[width=\linewidth]{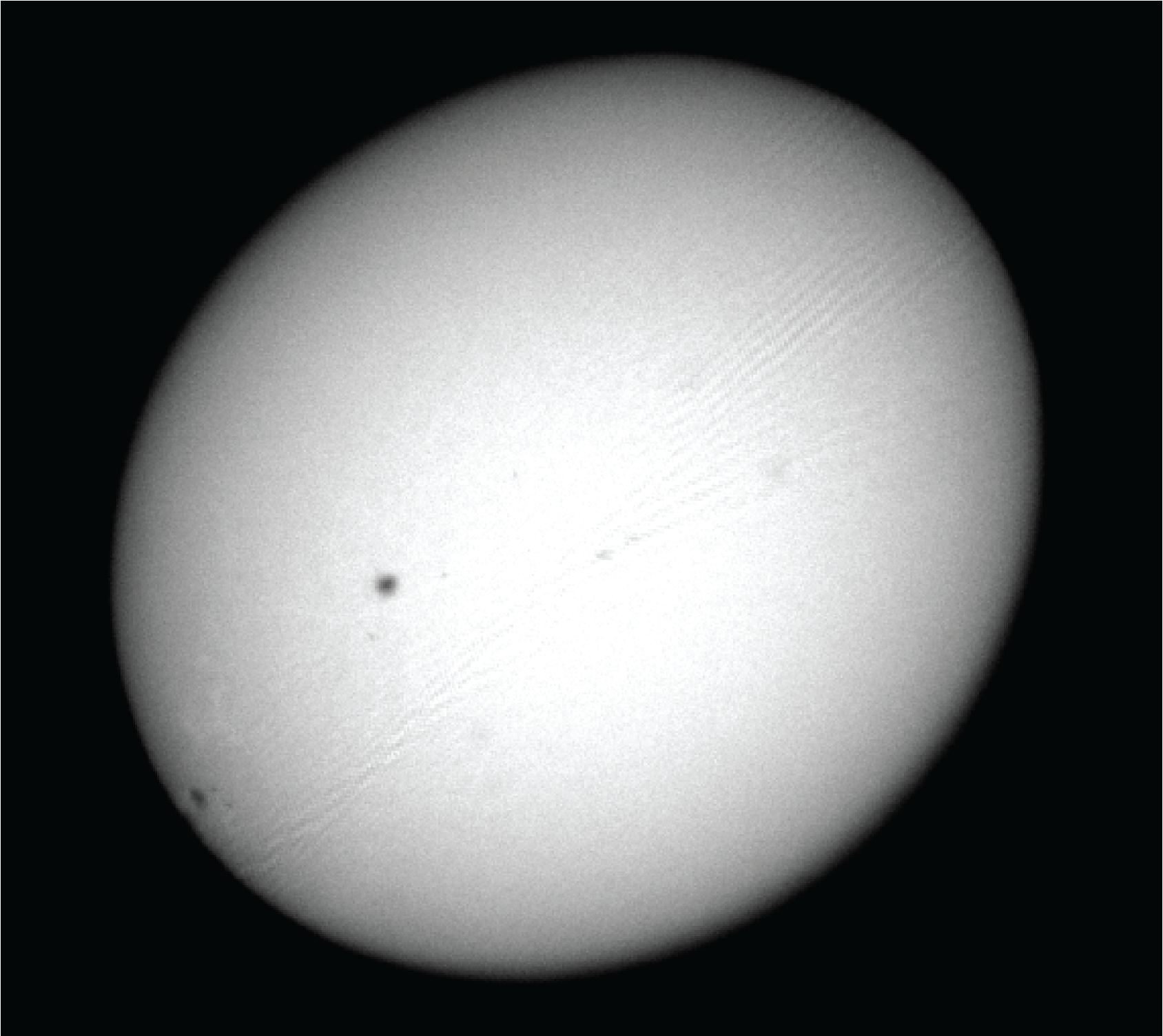} 
        \caption{}
        \label{fig:sun}
    \end{subfigure}
    \hfill
    \begin{subfigure}[t]{0.45\textwidth}
        \centering
        \includegraphics[width=\linewidth]{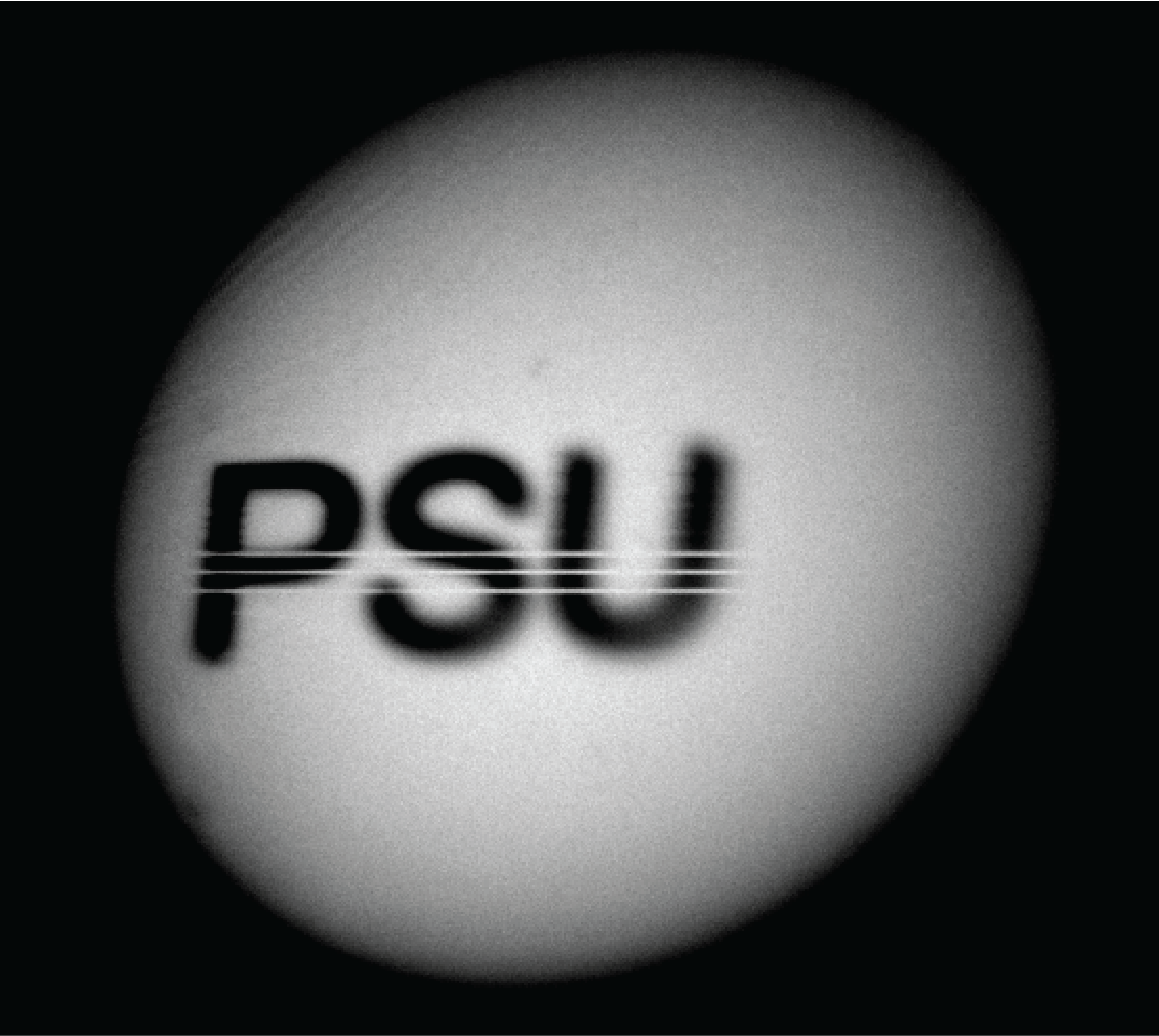} 
        \caption{} 
        \label{fig:dmd_sun}
    \end{subfigure}
    \caption{Images captured from the imaging channel of the MiraSOL prototype. (a) The DMD is commanding all micromirrors within the solar disk to the OFF state (directing light to the imaging channel), making sunspots clearly visible. (b) The DMD is dynamically masking a specific spatial region to spell out ``PSU" (Penn State University), directing that light away from the imaging channel. The visible horizontal banding is an artifact of the short camera exposures beating against the automated 60 Hz micromirror reset sequence discussed in Sec.~\ref{sec:dmd}.
    \label{fig:pics}
}
\end{figure*}

We conducted these initial observations with the prototype on the roof of Davey Laboratory at Penn State. Aided by a solar finder for telescope pointing, we captured images with the DMD in the all OFF state (directing all light to the imaging channel). As expected, the solar image exhibits some distortion due to the simple COTS optics used for this proof-of-concept. Fig.~\ref{fig:sun} shows sunspots clearly visible on the surface of the solar disk. Fig.~\ref{fig:dmd_sun} demonstrates the DMD dynamically modulating the light to create a dark region spelling ``PSU", the acronym for Penn State University. The horizontal banding present in the images is likely an artifact of the short camera exposures beating against the automated 60 Hz micromirror reset sequence discussed in Sec.~\ref{sec:dmd}. We plan to introduce an additional filter in front of the sensor to lengthen the exposure time and test this hypothesis. Ultimately, the knowledge gained from this prototype will directly inform the final optical design of MiraSOL.

\section{Conclusion}
In this proceeding, we have motivated the preliminary design requirements for MiraSOL, an EPRV solar feed that utilizes a digital micromirror device for spatially resolved solar observations. We built upon previous laboratory testing to further investigate key DMD properties, specifically identifying the 60 Hz flickering introduced by the manufacturer's evaluation board. We also presented a preliminary schematic illustrating the MiraSOL imaging and spectrometer channels, including a deployable fold mirror to bypass the DMD for traditional Sun-as-a-star observations. Finally, we demonstrated that a MiraSOL prototype built with COTS components can successfully image the Sun and produce dynamic, patterned masks across the solar disk. This validates the core concept of using a DMD for solar spatial modulation and serves as a vital pathfinder. The lessons learned from this prototype will be incorporated into the final optomechanical design of the MiraSOL instrument.

\acknowledgments   

This material is based upon work supported by the National Science Foundation Graduate Research Fellowship Program under Grant No. DGE1255832. Any opinions, findings, and conclusions or recommendations expressed in this material are those of the author(s) and do not necessarily reflect the views of the National Science Foundation. This work was partially supported by funding from the Center for Exoplanets and Habitable Worlds. The Center for Exoplanets and Habitable Worlds is supported by the Pennsylvania State University and the Eberly College of Science. Gemini AI (Google) was used for technical editing and structural organization assistance.

We acknowledge help from Farangiz Kholmatova, who, assisted in prototype alignment and rooftop observations. Elizabeth Gonzalez and Jessica Libby-Roberts gave input into the science and measurement requirements. Steve Smee shared his experience and insight into DMD uses in astronomy and help in trying to deconstruct the package to explore window replacement.

% References
\bibliography{ref} % bibliography data in report.bib
\bibliographystyle{spiebib} % makes bibtex use spiebib.bst

\end{document}